\documentclass{article}

\usepackage{array}
\usepackage{enumitem}

\usepackage{PRIMEarxiv}

\usepackage[utf8]{inputenc} 
\usepackage[T1]{fontenc}    
\usepackage{hyperref}       
\usepackage{url}            
\usepackage{booktabs}       
\usepackage{amsfonts}       
\usepackage{nicefrac}       
\usepackage{microtype}      
\usepackage{lipsum}
\usepackage{fancyhdr}       
\usepackage{graphicx}       
\graphicspath{{media/}}     

\title{Evaluating the Impact of Adversarial Traffic Patterns on VANET Communication Using Veins Simulation
\thanks{\textit{\underline{Citation}}: 
\textbf{Authors. Title. Pages.... DOI:000000/11111.}} 
}

\author{Henry Agyapong \\
  \texttt{Texas A\&M University} \\
  San Antonio \\
  \texttt{hagya01@jaguar.tamu.edu} \\
}
\begin{document}
\maketitle

\begin{abstract}
Vehicular Ad Hoc Networks (VANETs) are a key component of intelligent transportation systems, enabling real-time communication between vehicles. However, their open and dynamic nature makes them highly vulnerable to adversarial behaviors that can disrupt communication reliability. This paper investigates the impact of adversarial traffic patterns on VANET performance using the Veins simulation framework integrated with OMNeT++ and SUMO.

We design and evaluate multiple adversarial scenarios, including message flooding, false information dissemination, and coordinated congestion attacks, under varying traffic densities and mobility conditions. The study measures key performance metrics such as packet delivery ratio (PDR), end-to-end delay, and network throughput.

Experimental results show that adversarial traffic can reduce PDR by up to 96.55\%, with message flooding at low density producing a throughput reduction of 27.89\%, and significantly degrade overall network efficiency. The findings highlight critical vulnerabilities in VANET communication and provide insights into designing more resilient and secure vehicular networks.
\end{abstract}
\keywords{VANET, Adversarial Networks, Network Security, Veins, OMNeT++, SUMO, Packet Delivery Ratio, Intelligent Transportation Systems}
\section{Introduction}
Vehicular Ad Hoc Networks (VANETs) enable communication among vehicles and roadside units, supporting applications such as traffic management, collision avoidance, and autonomous driving. Despite their benefits, VANETs are inherently vulnerable due to their decentralized architecture, high mobility, and reliance on wireless communication.

One of the most critical yet underexplored threats is adversarial traffic behavior, where malicious or compromised vehicles inject disruptive communication patterns into the network. These behaviors can include message flooding, dissemination of false data, or coordinated attacks that degrade network performance.

While prior research has explored VANET routing and performance under normal conditions, limited work has systematically evaluated the impact of adversarial traffic patterns under realistic mobility scenarios.

\subsection{Research Objectives}

This paper aims to:
\begin{itemize}
  \item Analyze how adversarial vehicle behaviors affect VANET communication performance
\item Quantify the degradation in key network metrics under different attack scenarios
\item Provide insights into vulnerabilities and resilience limitations in VANET systems
  
\end{itemize}

\subsection{Contributions}

\begin{itemize}
  \item A simulation-based framework for modeling adversarial VANET behaviors using Veins
\item Design of multiple adversarial traffic scenarios (flooding, false data, coordinated attacks)
\item Quantitative evaluation of network degradation under varying densities
\item Insights into the security and robustness of VANET communication

\end{itemize}

A simulation-based framework for modeling adversarial VANET behaviors using Veins
Design of multiple adversarial traffic scenarios (flooding, false data, coordinated attacks)
Quantitative evaluation of network degradation under varying densities
Insights into the security and robustness of VANET communication

\subsection{ Adversarial Traffic Models}

We design three adversarial behaviors:

\begin{itemize}
  \item Message Flooding Attack: Malicious vehicles generate excessive messages
Objective: overload communication channels
\item False Information Dissemination: Nodes inject incorrect traffic or safety messages
Objective: disrupt network reliability
\item Coordinated Congestion Attack: Multiple adversarial nodes synchronize transmissions
Objective: maximize packet collisions and delay

\end{itemize}
\section{RELATED WORK}
Existing research on VANETs has primarily focused on routing protocols, mobility models, and performance optimization under benign conditions. Simulation frameworks such as Veins (integrating OMNeT++ and SUMO) have been
widely used to study realistic vehicular communication scenarios.
\\\\
Several studies have evaluated VANET performance metrics such as packet delivery ratio, delay, and throughput under
varying traffic densities. However, these studies often assume cooperative and non-malicious environments.
\\\\
Recent work in network security highlights vulnerabilities in wireless ad hoc networks, including denial-of-service
attacks, message tampering, and Sybil attacks. In VANETs, adversarial behavior can be particularly impactful due to
the real-time nature of communication.
\\\\
There is limited empirical evaluation of how adversarial traffic patterns—especially coordinated and mobility-aware
attacks—affect VANET performance in realistic simulations.\\\\
This paper aims to bridge the research gap by simulating and analyzing the three adversarial traffic behaviors discussed above namely: message flooding, disseminating false information, and congestion attacks within the context of different traffic density settings using the Veins simulation software.
\\\\
Veins is a simulation framework for the combination of vehicular traffic and network simulation. As observed by Sommer, German, \& Dressler (2011), they established that realistic modeling of vehicular traffic using SUMO contributes considerably in the performance measures such as communication range and delivery ratios of packets. In this study, additional investigations on the Veins simulation framework have been carried out based on hostile vehicular traffic concept that has not been explored in previous study.\\\\
The framework has been evaluated by Noori (2012) in his study regarding the usage of Veins framework in real-life vehicular ad-hoc network simulation in the city of Cologne, Germany. The study analyzed the impact of delivering beacon message in real life setting where each vehicle would broadcast beacon messages of size 256 bytes at 10Hz. Although Nori provides an interesting baseline for assessing the performance of VANETs under normal circumstances, no study has been done yet to evaluate the performance of VANETs under adversarial conditions. The present study seeks to fill this gap by systematically analyzing the effects of such behaviors on VANET performance using the same metrics employed in Nori and other studies as the basis for their analysis.\\\\

Di Pietro et al. (2014) offer an essential survey on security issues in different wireless ad hoc network models, with specific emphasis placed on VANETs. Selfish automobiles might try to clear their path ahead by generating false traffic alerts, or alternatively, generate false alerts for other automobiles to block the path of the police cars chasing the criminals. Such attacks could prove fatal to the police officers and civilians. The survey further points out Denial of Service attacks on the communications between vehicles and infrastructure as well as between the automobiles themselves. In this regard, the Distributed Denial of Service attack launched by multiple automobiles proves to be more severe because the attackers can communication.\\\\
More recently, a classification by Kaur et al. (2024) provides a detailed categorization of security vulnerabilities in VANETs in accordance with the layers of the communication protocols used. The authors show that various types of security vulnerabilities and attacks affect VANETs because of the nature of the networks and include mobility, dynamism, use of wireless communication technologies, and decentralization. Attacks at the physical layer, data link layer, network layer, transport layer, and application layer affect such security attributes as availability, integrity, and authenticity of the transmitted information. At the application layer, two attacks can be distinguished: the Illusion Attack and False Alert Generation, which involve the spread of false information about the condition of the roads in order to deceive vehicles or infrastructure nodes and cause misdirection, accidents, traffic jams, wrong routing, and hazards. At the network layer, Malicious Flooding, which involves flooding of the network with fake packets in order to use up network bandwidth and other network resources, resulting in network congestion, is discussed.\\
\section{EXPERIMENT GOALS AND DESIGN}
The goal of this experiment is to assess the effect of adversarial traffic conditions on VANET communication performance for different vehicle densities. The experiments will be specifically formulated to accomplish the objective by the following:
\begin{itemize}
\item Evaluate the communication performance of the VANET under normal traffic conditions in three vehicle densities; low, medium and high
\item Metrically quantify the effect of three different types of adversarial attacks on the network performance. They include flooding of messages, dissemination of false information and congestion
\item Determine weaknesses and limitations in terms of resistance against each of the attacks in different vehicle densities
\end{itemize}

\subsection{Simulation Environment}
The simulation architecture employs three main simulation tools: OMNeT++ used for discrete event network simulation, SUMO (Simulation of Urban Mobility) for realistic vehicle mobility, and the Veins 5.3.1 simulation framework as the bridge between the two.\\\\
OMNeT++ and SUMO interaction was achieved using the TraCI (Traffic Control Interface) protocol, which provides the ability to exchange the mobility and network information in real time during simulation process.\\\\
All the experiments performed utilized the same network of roads which was the case of Erlangen, available with the Veins framework which models a realistic urban road topology. The communication mode used by the vehicles in these experiments was that based on IEEE 802.11p, at the frequency of 5.9 GHz, which is the communication protocol used for VANETs. One RSU (Roadside Unit) was implemented in the simulation for facilitating Vehicle-to-Infrastructure communications along with V2V communications.

\subsection{Traffic Density Scenarios}
To assess the effects of adversarial actions under various network conditions, three traffic density setups are established:
\begin{itemize}
    \item \textbf{Low Density}: 10 vehicles are placed within the simulation zone, symbolizing sparse urban traffic with minimal vehicle interaction
    \item \textbf{Medium Density}: 25 vehicles are utilized, symbolizing average urban traffic with enhanced communication possibilities
    \item \textbf{High Density}: 50 vehicles are utilized, symbolizing crowded city traffic with significant node concentration and regular wireless communications
\end{itemize}
Each scenario runs for a simulation duration of 200 seconds. Vehicles enter the network gradually, with a 3-second interval between successive vehicles, allowing for a steady increase in traffic that aligns with authentic urban scenarios.\\

\subsubsection{Message Flooding Attack}
The message flooding attack replicates the case of denial of service in which malicious vehicles flood the network with too many messages and clog the communication channel. In this experiment, all the vehicles in the simulation become attackers by sending 50 extra WAVE Short Message (WSMs)  on top of one regular message. Hence, each vehicle will be sending 51 messages in total. The resulting packet number will be 510, 1275, and 2550 in the low density, medium density, and high density simulations respectively. The main goal of this attack is to overload the communication channel.

\begin{figure}[ht]
    \centering
    \includegraphics[width=0.75\linewidth]{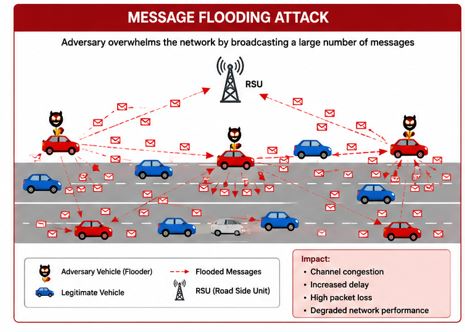}
    \caption{Message Flooding Attack}
\end{figure}

\subsubsection{False Information Dissemination Attack}
The false information dissemination attack attempts to reproduce a situation where hacked vehicles will generate wrong information about traffic and safety within the communication network. One-third of vehicles, whose node identifiers are multiples of three, will be used as malicious actors in the network. The attackers are programmed to generate messages with wrong road identifier information, and not with the road identifier of their own. False data in the form of road identifiers, false emergency messages, and false congestion warning messages are generated. To ensure that the simulation does not attempt to validate the false road information, normal vehicles are instructed to ignore the messages with such identifiers.\\ [9em]

\begin{figure}[ht]
    \centering
    \includegraphics[width=0.75\linewidth]{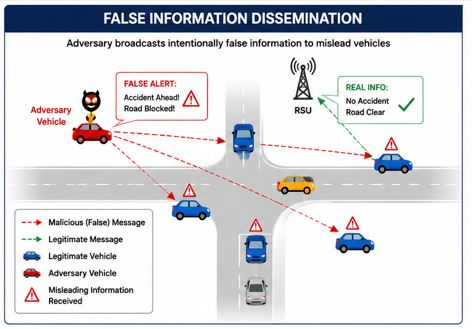}
    \caption{False Information Dissemination Attack}
\end{figure}

\subsubsection{Coordinated Congestion Attack}
The coordinated congestion attack builds upon the idea of flooding attack by adding a synchronization factor to the transmission pattern. In the simulation model, one out of every two vehicles is considered to be an attacker, creating a 50\% attacker ratio. The attacking nodes transmit three packets in every burst for a maximum of 30 transmissions, thus transmitting an additional 90 packets as compared to normal behavior. While in the flooding attack the transmissions happen continuously, coordination in this type of attack aims to create maximum channel occupancy for achieving more collisions and congestions.

\begin{figure}[ht]
    \centering
    \includegraphics[width=0.75\linewidth]{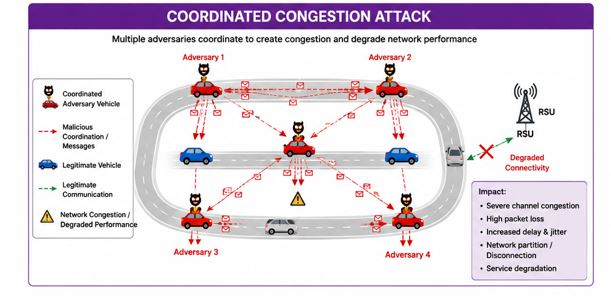}
    \caption{Coordinated Congestion Attack}
\end{figure}

\subsection{Performance Metric}
The following performance metrics have been applied for assessing and comparing the communication performances of VANETs in all of the baseline and adversarial cases presented in this paper. Each metric was chosen for capturing an unique aspect of the network’s functionality. This will provide a thorough understanding of the effect of different types of adversary traffic on the performance of communications.

\begin{itemize}
    \item \textbf{Packet Delivery Ratio (PDR):} Packet Delivery Ratio is one of the major performance metrics used in this paper.  Packet Delivery Ratio measures the proportion of successfully received broadcast packets relative to the total number of packet delivery attempts at the MAC layer, expressed as a percentage. For this paper, the MAC layer metric is considered because the attack affects the wireless channel. The use of Veins broadcast model also leads to an inconsistency in the number of packet receptions at the application layer. PDR is defined as:\\

PDR (\%) =(Total Received Broadcasts/(Total Received Broadcasts + Total Lost Packets)) × 100\\
    \item \textbf{Network Throughput :} Network throughput refers to the amount of information sent through the network effectively within a certain period of time, measured in bits per second (bps), and kilobits per second (Kbps). Throughput indicates the capacity of data transfer in the VANET network under normal circumstances, as well as when attacked by hackers.\\

Throughput (bps) = (Number of Received Broadcast Packets × Packet Size in Bits) / Simulation Duration (Seconds)\\\\
It should be noted that all packets transmitted in this experiment are equal to 1,104 bits, and the duration of simulations is 200 seconds.\\

    \item \textbf{Total Packet Loss:} Total number of packets which could not be delivered because of channel interferences or because of the bad signal-to-noise ratios or simultaneous transmissions. This is automatically calculated by Veins MAC Layer components and saved in .sca (scalar result files).
\end{itemize}

\section{EXPERIMENT ANALYSIS AND DISCUSSION}
\subsection{Baseline Performance}
Prior to introducing adversarial behaviors, baseline experiments were conducted across all three traffic density configurations to establish reference performance values for normal VANET operation. 

\begin{table}[h]
{\centering
\begin{tabular}{|p{1.5cm}|p{1.5cm}|p{1.5cm}|p{1.5cm}|p{1.5cm}|p{1.5cm}|p{1.5cm}|}
\hline
\multicolumn{1}{|c|}{\textbf{ Traffic Density}} & 
\multicolumn{1}{c|}{\textbf{Number of Vehicles}} & 
\multicolumn{1}{c|}{\textbf{Packet Sent}} & 
\multicolumn{1}{c|}{\textbf{Broadcast Received}} &
\multicolumn{1}{c|}{\textbf{Packets Lost}} &
\multicolumn{1}{c|}{\textbf{PDR(\%)}} &
\multicolumn{1}{c|}{\parbox{2cm}{\centering\textbf{Throughput} \\\textbf{(bps)}}} \\

\hline

\textbf{Low} &
10 &
10 &
150 &
0 &
100 &
1048.80 \\

\hline
\textbf{Medium} &
25 &
25 &
549 &
19 &
96.65 &
3030.48 \\

\hline

\textbf{High} &
50 &
31 &
728 &
51 &
93.45 &
4018.56 \\

\hline

\end{tabular}
\par}
\caption{Baseline Results}
\label{tab:related_work}
\end{table}

Under baseline conditions, the low-density scenario resulted in an optimal Packet Delivery Ratio of 100 percent with all 10 packets transferred successfully and no packet losses recorded along with an improved network throughput of 1,048.80 bps.\\\\
As traffic density was raised to medium, the PDR dropped to 96.65 percent due to 19 packet losses among the 25 total transferred packets with a further increase in network throughput to 3,030.48 bps. This happened due to the increased competition among more vehicles using the same wireless network channel, which led to some occasional collisions between packets and caused packet losses at the MAC layer.\\\\
The highest density baseline had a low PDR of 93.45\% due to the loss of 51 packets, although the throughput rose even further to 4,018.56 bps. It is important to observe that out of the 50 vehicles placed in the simulation, only 31 were able to transmit packets because the rest joined the network after the requisite 10 seconds stopping time was achieved for triggering a packet transmission.

\subsection{Message Flooding Attack Analysis}
The message flooding attack was implemented by configuring all vehicles to transmit 51 packets each — one standard message plus 50 additional flood packets — resulting in a 51-fold increase in network traffic compared to the baseline. 
\begin{table}[h]
{\centering
\begin{tabular}{|p{1.5cm}|p{1.5cm}|p{1.5cm}|p{1.5cm}|p{1.5cm}|p{1.5cm}|p{1.5cm}|}
\hline
\multicolumn{1}{|c|}{\textbf{ Traffic Density}} & 
\multicolumn{1}{c|}{\textbf{Number of Vehicles}} & 
\multicolumn{1}{c|}{\textbf{Packet Sent}} & 
\multicolumn{1}{c|}{\textbf{Broadcast Received}} &
\multicolumn{1}{c|}{\textbf{Packets Lost}} &
\multicolumn{1}{c|}{\textbf{PDR(\%)}} &
\multicolumn{1}{c|}{\parbox{2cm}{\centering\textbf{Throughput} \\\textbf{(bps)}}} \\

\hline

\textbf{Low} &
10 &
510 &
137 &
3,832 &
3.45 &
756.24 \\

\hline
\textbf{Medium} &
25 &
1,275 &
877 &
15,045 &
5.51 &
4,841.04 \\

\hline

\textbf{High} &
50 &
2,550 &
1,889 &
44,605 &
4.06 &
10,427.28 \\

\hline

\end{tabular}
\par}
\caption{Flooding Attack Results }
\label{tab:related_work}
\end{table}

Message flooding attack caused a major degradation in network performance. The PDR value was reduced drastically from a 100\% perfect rate to only 3.45\%, which constitutes a decrease of 96.55 percentage points in the low density case.\\\\
The total lost packets were increased dramatically to 3,832 from zero in the case of the baseline test under the flooding attack, indicating that even a few flooding nodes can overload a sparsely populated VANET.\\\\
The results for medium density flooding indicated a PDR value of 5.51\%, but there was a huge number of packets that got lost (15,045), which was over 790 times more than in the baseline test (19 packets). There was a similar case for the high density scenario, in which the PDR value was only 4.06\%, but 44,605 packets were lost overall. Overall, the consistently low PDR values (less than 6\%) across the three densities prove the efficacy of the flooding attack irrespective of the density, thereby rendering communication almost impossible because the channel gets saturated.\\\\
The sudden spike in packet loss can be attributed to mainly two processes. The first is Signal-to-Noise plus Interference Ratio (SNIR) degradation caused by the saturation of the channel due to multiple transmission events, whereby the signal strength deteriorates beyond the reception threshold limit of most of the packets. The second is the occurrence of transmission/reception conflicts at the MAC layer, which increases sharply because every node tries transmitting at maximum capacity. It can be concluded from these results that the flooding of messages poses serious DoS threats to VANET communications and can lead to a network loss of over 90\%.

\subsection{False Information Dissemination Attack Analysis}
The false information dissemination attack was implemented by designating approximately one-third of vehicles as attackers, each transmitting fabricated road identifiers, false emergency alerts, and false congestion warnings in place of legitimate traffic data. Table 3 presents the false information attack results.
\begin{table}[h]
{\centering
\begin{tabular}{|p{1.5cm}|p{1.5cm}|p{1.5cm}|p{1.5cm}|p{1.5cm}|p{1.5cm}|p{1.5cm}|}
\hline
\multicolumn{1}{|c|}{\textbf{ Traffic Density}} & 
\multicolumn{1}{c|}{\textbf{Number of Vehicles}} & 
\multicolumn{1}{c|}{\textbf{Packet Sent}} & 
\multicolumn{1}{c|}{\textbf{Broadcast Received}} &
\multicolumn{1}{c|}{\textbf{Packets Lost}} &
\multicolumn{1}{c|}{\textbf{PDR(\%)}} &
\multicolumn{1}{c|}{\parbox{2cm}{\centering\textbf{Throughput} \\\textbf{(bps)}}} \\

\hline

\textbf{Low} &
10 &
2,257 &
1,675 &
8,778 &
16.02 &
9,246.00 \\

\hline
\textbf{Medium} &
25 &
4,987 &
4,863 &
37,688 &
11.43 &
26,843.76 \\

\hline

\textbf{High} &
50 &
8,314 &
11,261 &
90,115 &
11.11 &
62,160.72 \\

\hline

\end{tabular}
\par}
\caption{False Information Results }
\label{tab:related_work}
\end{table}\\
The false information attack created a unique and more complicated pattern of degradation compared to the flooding attack. The overall number of packets sent was significantly greater in all cases of density – 2,257 packets sent in the case of low density, 4,987 packets sent in the case of medium density, and 8,314 packets sent in the case of high density.\\\\
In all densities of false information, the percentage delivery ratio (PDR) varied between 11.11\% and 16.02\%, which is quite an improvement compared to the flooding attack but is still a drastic decrease in performance. The low-density case had the greatest percentage delivery ratio (PDR) of 16.02\%, alongside a throughput of 9,246.00 bps, while in medium and high-density cases, the value of PDR was almost the same at about 11\%.\\\\
The unique feature of this type of attack lies in the double impact on the reliability of communication networks. First, due to the high volume of traffic sent by the malicious entity, packets are being lost similarly to a flooding attack, which is proved by the high packet losses ranging from 83.98\% to 88.89\%. However, at the application layer, if some packets are still transmitted correctly, they include incorrect data, i.e., wrong roads' identifiers, fake emergency messages, and erroneous notifications about congestion. Therefore, this type of malicious activity has two negative impacts on VANETs since both quantity and quality of information are being compromised.

\subsection{Coordinated Congestion Attack Analysis}
The coordinated congestion attack was implemented by configuring every second vehicle as an attacker, each transmitting burst packets in coordinated cycles designed to maximize simultaneous channel occupancy. 
\begin{table}[h]
{\centering
\begin{tabular}{|p{1.5cm}|p{1.5cm}|p{1.5cm}|p{1.5cm}|p{1.5cm}|p{1.5cm}|p{1.5cm}|}
\hline
\multicolumn{1}{|c|}{\textbf{ Traffic Density}} & 
\multicolumn{1}{c|}{\textbf{Number of Vehicles}} & 
\multicolumn{1}{c|}{\textbf{Packet Sent}} & 
\multicolumn{1}{c|}{\textbf{Broadcast Received}} &
\multicolumn{1}{c|}{\textbf{Packets Lost}} &
\multicolumn{1}{c|}{\textbf{PDR(\%)}} &
\multicolumn{1}{c|}{\parbox{2cm}{\centering\textbf{Throughput} \\\textbf{(bps)}}} \\

\hline

\textbf{Low} &
10 &
3,142 &
3,187 &
12,784 &
19.95 &
17,592.24 \\

\hline
\textbf{Medium} &
25 &
7,222 &
8,275 &
67,783 &
10.88 &
45,678.00 \\

\hline

\textbf{High} &
50 &
12,925 &
22,798 &
166,848 &
12.02 &
125,844.96 \\

\hline

\end{tabular}
\par}
\caption{Coordinated Congestion Attack Results }
\label{tab:related_work}
\end{table}\\
The coordinated congestion attack produced the highest traffic volume of 3,142, 7,222, and 12,925 packets respectively for low, medium, and high density networks. Nevertheless, even at such an elevated level of traffic, the values of the PDR equal to 19.95\%, 10.88\%, and 12.02\% are relatively higher than those experienced in the flooding attacks implying that the coordinated attacks generate interference slightly differently from flooding attacks considering the burst-based nature of the traffic generation process in these attacks.\\\\
The low density coordinated congestion attack resulted in a 19.95\% PDR — the highest for all the three densities – along with a throughput of 17,592.24bps. The result is in line with sparse density networks where although the coordinated bursts were intense, there was enough time between bursts during which packets could actually be transmitted successfully. However, when the number of nodes increased, the PDR tended to fall into the range of 11\%-12\% for medium and high density networks respectively due to the coordinated bursts adding up to the already high number of nodes.\\\\
The combined result of the congestion attack recorded the greatest amount of packet losses compared to any other attack conducted in the study, with an amount of 166,848 packets lost and 125,844.96 bps throughput mainly because of the high number of attack traffic being sent into the network. The difference between this result and those of flooding attack and the false information attack is significant because of the greater absolute number of packets lost, especially due to the 50 percent attacker ratio and the burst effect of packets three times per attacker.

\section{Comparative Analysis}
Several remarkable features concerning the specifics of hostile activity in relation to VANETs can be singled out based on the results of the comparison made. First of all, the message flooding attack produced the lowest rate of PDR, with the figure being below 6\% for each variation in density. It goes without saying that such an attack proves to be the most efficient way of reducing the reliability of the channel. The false information attack and the coordinated congestion attack resulted in a PDR within 10\% and 20\%, respectively; thus, it could be stated that these attacks constitute relatively high but somewhat smaller disruptions to the process.\\\\
In all three variants of attacks, there was a decrease in the value of PDR from the case of low traffic density to medium traffic density, while the figure stayed constant from the point of medium traffic density to high traffic density. Thus, despite the fact that high density channels provide better conditions for communication, they produce a greater impact of hostile attacks.\\\\
One noteworthy conclusion from the throughput analysis is that the raw throughput figures increase in attack situations at medium and high densities, due to the significantly larger amount of assault-generated traffic received. But, the PDR values indicate that the ratio of successfully delivered legitimate packets drops to less than 6\% in the presence of flooding. This shows that throughput alone is not a reliable measure of network reliability in the presence of an adversary and it should be used together with PDR and packet loss. \\\\
Overall, the results indicate that VANET communication is very vulnerable to all three malicious traffic patterns considered in this work. The mildest attack, coordinated congestion at low density, reduced PDR from 100\% to 19.95\%, an 80 percentage point loss in network reliability. The results demonstrate important security shortcomings in current VANET communication protocols and point to the necessity of detection and mitigation systems that can detect and neutralize adversarial behaviors in the context of vehicular networks.

\section{Conclusion and Future Work}
\subsection{Conclusion}
In this paper, we have analyze the impact of following adversarial attacks on the effectiveness of the process of communication in VANET: message flooding attack, false information dissemination and coordinated congestion. These adversarial attacks were performed in three different levels of traffic density: low, medium, and high. We estimated their performance on the basis of Packet Delivery Rate, packets loss, and network throughput.\\\\
This experiment has demonstrated that all these adversarial attacks considerably reduce the reliability of the network. Message flooding is the most dangerous type of attack in terms of communication effectiveness causing decreasing of Packet Delivery Rate by 96.55\% and throughput by 27.89\% at low traffic density level. Also, this adversarial attack completely stops the communication in the network. The distribution of false information decreases the PDR by 83.98\% and corrupts the information transmitted in the messages. The coordinated congestion has the maximum packet loss under the condition of high traffic density causing the loss of 166,848 packets.\\\\
To sum up, the obtained results emphasize the vulnerability of modern VANETs to malicious traffic types. Namely, PDR is significantly lowered (from 80\% up to 96.55\%). The findings obtained demonstrate that the considered forms of malicious traffic may seriously disrupt communication processes within VANETs.

\subsection{Future Work}
A number of avenues for further research based on this study’s results and limitations present themselves.
Hybrid attack strategies using multiple simultaneous adversarial behaviors need to be studied to determine their compounded effect, as real-life adversaries are unlikely to use only one kind of adversarial behavior at once. The simulation framework created by this study is a solid baseline that makes it possible to evaluate hybrid strategies’ efficiency in detail.
Further research into developing and analyzing techniques for detecting and counteracting the adversarial strategies discovered in this study, given their high risk to safety when used on VANETs, is a valuable area of further research.
Finally, evaluating the impact of adversarial behavior in large-scale and geographically diverse networks is crucial to finding out whether the patterns of attack severity discovered here apply universally to a variety of road network topologies.

\nocite{*}
\bibliographystyle{unsrt}
\bibliography{references}

\end{document}